\documentclass[twocolumn]{aastex631}
\usepackage{hyperref}

\shorttitle{The BH spin tilt in GW200115}
\shortauthors{Xing-Jiang Zhu}

\begin{document}

\title{On the large apparent black hole spin-orbit misalignment angle in GW200115}

\email{zhuxj@bnu.edu.cn}

\author[0000-0001-7049-6468]{Xing-Jiang Zhu}
\affil{Advanced Institute of Natural Sciences, Beijing Normal University at Zhuhai 519087, China}

\begin{abstract}
GW200115 is one of the first two confidently detected gravitational-wave events of neutron star-black hole mergers.
An interesting property of this merger is that the black hole, if spinning rapidly, has its spin axis negatively aligned (with a misalignment angle $> 90^{\circ}$) with the binary orbital angular momentum vector.
Although such a large spin-orbit misalignment angle naturally points toward a dynamical origin, the measured neutron star-black hole merger rate exceeds theoretical predictions of the dynamical formation channel.
In the canonical isolated binary formation scenario, the immediate progenitor of GW200115 is likely to be a binary consisting of a black hole and a helium star, with the latter forming a neutron star during a supernova explosion.
Since the black hole is generally expected to spin along the pre-supernova binary orbital angular momentum axis, a large neutron star natal kick is required to produce the observed misalignment angle.
Using simple kinematic arguments, we find that a misalignment angle $> 90^{\circ}$ in GW200115-like systems implies a kick velocity $\sim 600\, \text{km/s}$ and a kick direction within $\approx 30 ^{\circ}$ of the pre-supernova orbital plane.
We discuss different interpretations of the large apparent black hole spin-orbit misalignment angle, including a non-spinning black hole.
\end{abstract}


\section{Introduction}
Recently, \citet{LVKnsbh21} reported the observation of gravitational waves from two neutron star-black hole (NS-BH) coalescence events during the third observing run of the LIGO \citep{aLIGO} and Virgo \citep{aVirgo} detectors.
These two events, named GW200105 and GW200115, were the first confidently detected NS-BH binaries via any observational means.
They represent a novel subclass of compact binaries, providing renewed opportunities to study the formation of BHs, NSs and NS-BH systems.

From the gravitational-wave data analysis, the BH mass\footnote{The mass measurements quoted here are median and 90\% credible intervals, under the low-NS-spin prior. Throughout the paper, the BH refers to the primary merging object rather than the merger remnant.} is measured to be $8.9_{-1.3}^{+1.1}\ M_{\odot}$ and $5.9_{-2.1}^{+1.4}\ M_{\odot}$, and the NS mass to be $1.9_{-0.2}^{+0.2}\ M_{\odot}$ and $1.4_{-0.6}^{+0.2}\ M_{\odot}$, for GW2001015 and GW200115, respectively.
For such asymmetrical mass ratios, the measurable spin effects are dominated by the contribution from the more massive BHs.
For GW200105, the magnitude of the BH spin is constrained to be less than $0.23$ at the 90\% credible level.
For GW200115, the BH spin is found to be negatively aligned with the binary orbital angular momentum vector, i.e., with a spin-orbit misaligned angle greater than $90^{\circ}$.

In this work, we focus on the BH spin measurement of GW200115 and address specifically the astrophysical implications of a negatively aligned BH spin in GW200115-like systems.
In Section \ref{sec:spin}, we examine the observational evidence for BH spin in GW200115.
In Section \ref{sec:constrain}, we derive constraints on NS natal kicks assuming that 1) GW200115 contains a rapidly spinning BH, and 2) the BH is the first-formed compact object in the binary, which is formed through the isolated binary evolution channel.
In Section \ref{sec:discuss}, we relax those assumptions and expand on the discussion about the origin of GW2001115 (as well as GW200105) and their astrophysical interpretations.
Finally, we present concluding remarks in Section \ref{sec:conclude}.

\section{Is the black hole in GW200115 spinning?}
\label{sec:spin}

Here, we reassess the observational evidence for BH spin in GW200115.
The dimensionless spin magnitude $\chi_1$ (defined to be between 0 and 1) is measured to be $0.31_{-0.29}^{+0.52}$ ($0.33_{-0.29}^{+0.48}$) assuming the low $\chi_{2}<0.05$ (high $\chi_{2}<0.99$) NS-spin prior.
As mentioned in \citep{LVKnsbh21}, the BH spin magnitude $\chi_1$ of GW200115 is consistent with zero.
A quantitative measure of the statistical support for BH spin is the Bayes factor between a spinning-BH model and a non-spinning model.
We compute this Bayes factor with the publicly available posterior samples\footnote{\url{https://dcc.ligo.org/LIGO-P2100143/public}} for $\chi_1$, using the Savage-Dickey density ratio \citep[see, e.g., Appendix B of][]{Katerina14_spin}.
We find a Bayes factor in favor of BH spin of $\mathcal{B}=0.5\, (0.7)$ for the low (high) NS-spin prior.
Adjusting the BH prior range ($\chi_{1}<0.99$), e.g., adopting the upper limit on $\chi_1$ for the other NS-BH merger GW200105 in a new prior, makes little impact on the Bayes factors.
Therefore, the gravitational-wave data show no evidence for or against a rapid BH spin for GW200115.
If the BH is indeed rapidly spinning, its spin-orbit misalignment angle should be greater than $90^{\circ}$ with $\sim 90\%$ credibility \citep{LVKnsbh21}.

\section{Possible constraints on neutron star natal kicks}
\label{sec:constrain}

In the canonical isolated binary evolution channel, the progenitor of GW200115 is a binary of two massive stars.
It is generally expected that the more massive star in this binary ends its life first and forms a BH \citep[e.g.,][]{Fryer99,Floor21a}.
In this work, we only consider the \textit{immediate} progenitor of GW200115: a binary composed of a BH and a helium star that is about to go supernova (SN) and make a NS.
Our goal is to use the measured BH spin misalignment angle of GW200115 to constrain the SN kicks associated with the birth of the NS.
Since there is inconclusive evidence for BH spin in GW200115, our constraints should be taken as plausible rather than definitive.
For this reason, we simplify our calculations by fixing the BH/NS masses of GW200115 to their median posterior values.

The effect of SN kicks on the dynamics of a binary system is well studied in the literature \citep[e.g.,][]{Hills83,Bailes88,Brandt95,Kalogera00_SpinTilt,Hurley02}.
We follow the formalism described in section 2 of \citet{Brandt95} and define the following parameters: the BH mass $m_1$, the mass of the carbon-oxygen (CO) core $m_{\rm CO}$ of the helium star, the pre-SN orbital separation $a_0$, the NS mass $m_2$, the kick velocity $v_{\rm kick}$ and kick angles $(\theta,\ \phi)$, and the BH spin tilt angle $\alpha$ after the SN explosion.

\begin{figure}[ht]
\begin{center}
  \includegraphics[width=0.46\textwidth]{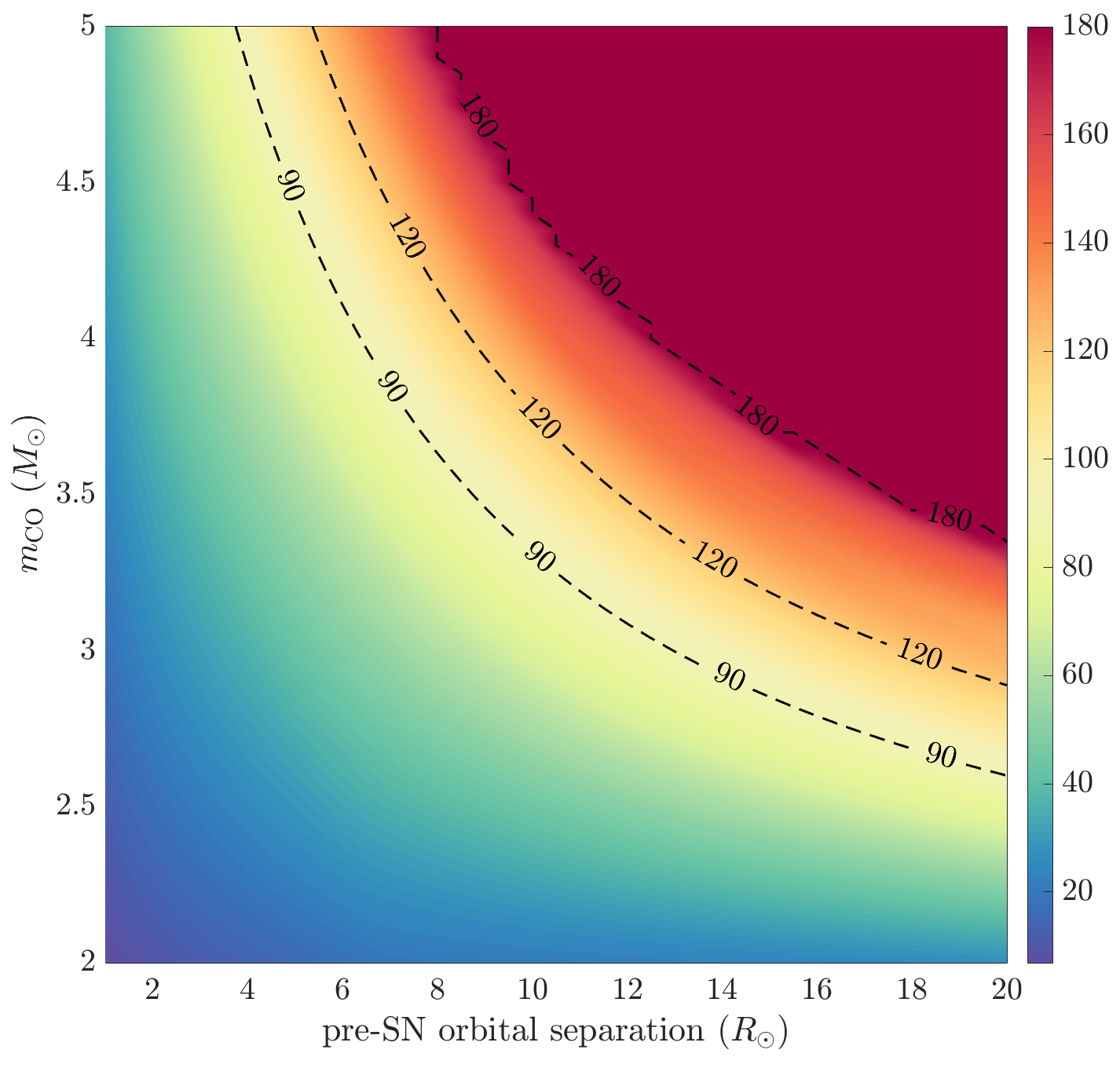}
  \caption{The typical BH spin tilt angle (in degrees, shown as coded color map) as a function of the pre-SN orbital separation and the mass of the carbon-oxygen (CO) core that gives rise to the NS. Note that the dark red region beyond the contour line of $180^{\circ}$ is unphysical: the binary gets disrupted during the SN or the post-SN binary is not capable of merging within the age of the Universe.
  \label{fig:BHtilt_contour}}
\end{center}
\end{figure}

We assume that the pre-SN binary orbit is circular and the BH spin is aligned with the orbital angular momentum prior to the SN explosion.
Following the simple recipe of \citet{MM20_receipe}, we adopt a linear relation between $v_{\rm kick}$ and the SN mass loss, namely, $v_{\rm kick} = \mu_{\rm kick} (m_{\rm CO}-m_{2})/m_{2}$, where $\mu_{\rm kick} =400\ {\rm km/s}$.
In order to produce a coalescing NS-BH binary, two constraints are applied: 1) the binary remains bound after the SN explosion; and 2) the binary is capable of merging within a Hubble time ($\approx 13.8$ Gyr).

In Figure \ref{fig:BHtilt_contour}, we show the typical (median) values of $\alpha$ for coalescing NS-BH binaries as a function of $m_{\rm CO}$ and $a_0$, assuming that $(m_{1}, m_{2})=(5.9, 1.4) M_{\odot}$ and the SN kicks are isotropically directed.
It can be seen that $\alpha > 90^{\circ}$ requires $m_{\rm CO} \gtrsim 2.5 M_{\odot}$ and $a_{0} \gtrsim 5 R_{\odot}$. However, they both cannot be too large, otherwise the binary gets disrupted by the SN kick or the resulting NS-BH system is not merging;
such a prohibited parameter space corresponds to the dark red region beyond the $\alpha=180^{\circ}$ contour line in Figure \ref{fig:BHtilt_contour}.

\begin{figure}[ht]
\begin{center}
  \includegraphics[width=0.46\textwidth]{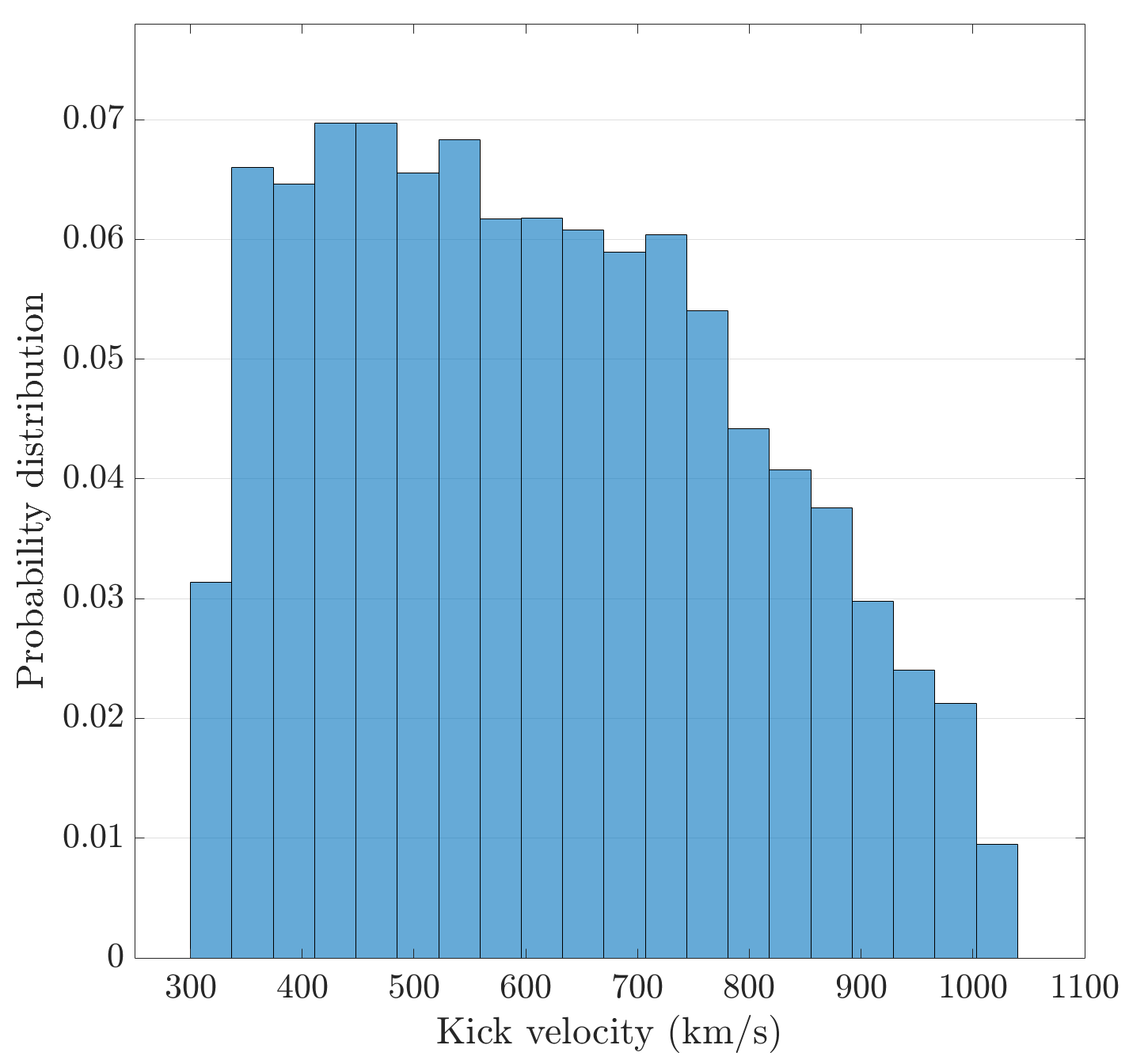}
  \caption{The probability distribution of the NS natal kick velocity in order to produce a GW200115-like NS-BH system with a BH spin tilt angle above $90^{\circ}$.
  \label{fig:kick_velocity}}
\end{center}
\end{figure}

\begin{figure}[ht]
\begin{center}
  \includegraphics[width=0.46\textwidth]{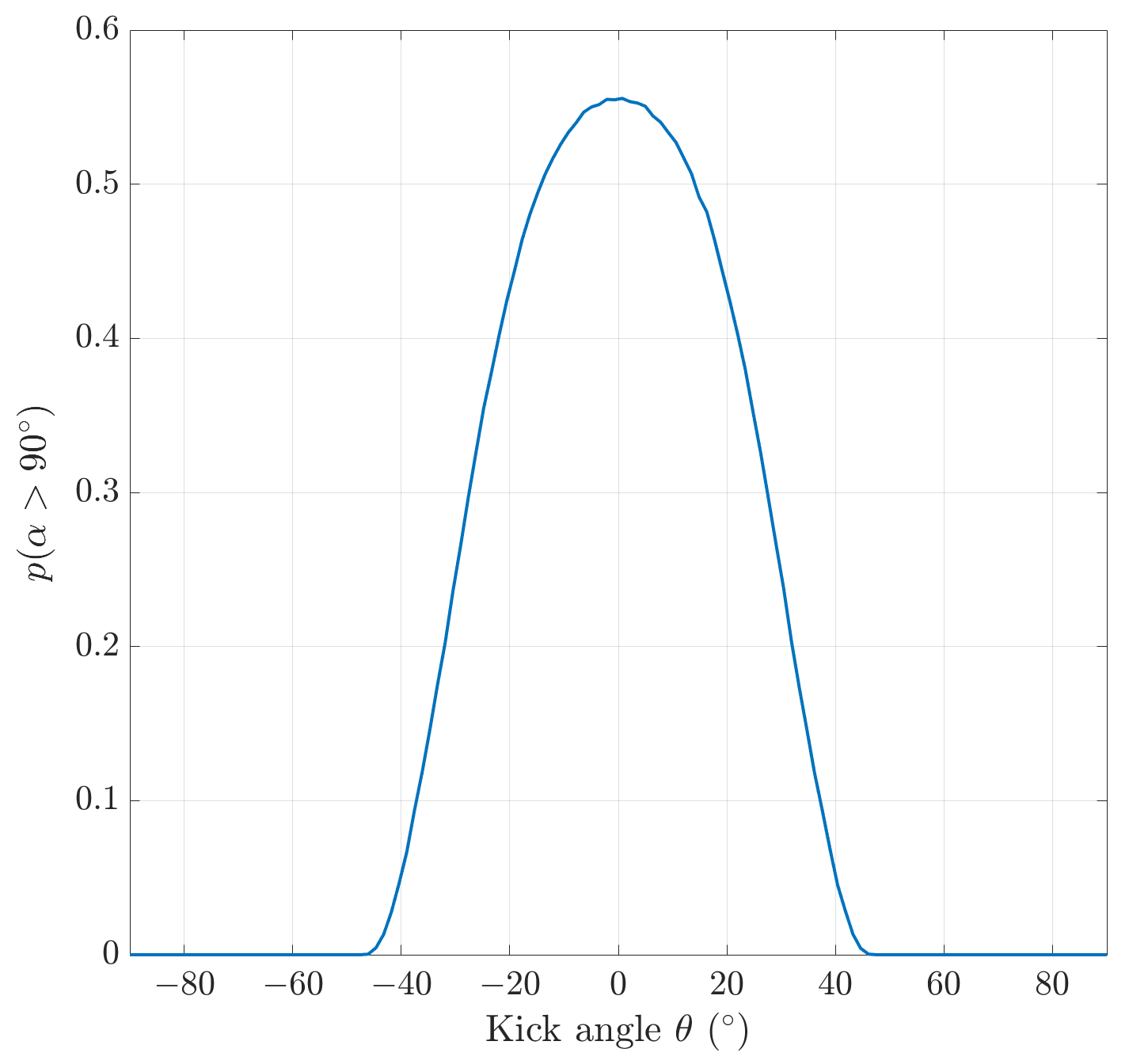}
  \caption{The probability that the BH spin tilt angle $\alpha$ is greater than $90^{\circ}$ as a function of the SN kick angle $\theta$, which is defined as the angle between the kick velocity vector and the pre-SN orbital plane.
  \label{fig:kick_angle}}
\end{center}
\end{figure}

Next, we simulate a population of NS-BH binaries with $(m_{1}, m_{2})=(5.9, 1.4) M_{\odot}$, $m_{\rm CO}$ uniformly distributed in $[2.5,5] M_{\odot}$, $a_{0}$ following a log-uniform distribution in $[5,30] R_{\odot}$.
Assuming isotropic kicks and requiring that 1) the binary remains bound after the SN, 2) the binary is capable of merging within a Hubble time, and 3) the post-SN BH spin is negatively aligned with the binary orbit ($\alpha > 90^{\circ}$), Figure \ref{fig:kick_velocity} shows the probability distribution of the kick velocity.
We find the kick velocity is $600_{-250}^{+340}\, \text{km/s}$ at the 90\% credible level.
This result is in agreement with a recent population synthesis study by \citet{Fragione21NSBH} who found that the detection of a GW200115-like event indicates large NS natal kicks: if the kick magnitudes are modelled as a Maxwellian distribution, the velocity dispersion $\sigma$ is greater than $150\, \text{km/s}$.
The result shown in Figure \ref{fig:kick_velocity} is also compatible with a Maxwellian distribution with $\sigma=265\, \text{km/s}$ proposed for pulsar birth velocity estimates based on proper motion measurements \citep{Hobbs05PSRkicks}.

Lastly, Figure \ref{fig:kick_angle} shows the probability of $\alpha > 90^{\circ}$ as a function of the kick angle $\theta$ between the kick velocity vector and the pre-SN orbital plane.
At first glance, the result might look counter-intuitive since in-plane kicks are usually associated with small spin tilt angles (while all else equal); $\theta=0$ leads to $\alpha=0$.
However, we are in a regime where the required kick velocity for $\alpha > 90^{\circ}$ is so high that the binary is easily disrupted, or, if not disrupted, its orbit may be too wide and thus the binary will not merge within a Hubble time.
The requirement for the binary to remain bound as well as to be capable of merging within a Hubble time leads to a preference for planar kicks (small $\theta$) against polar kicks (large $\theta$).
This is illustrated in Figure \ref{fig:kick_tilt} in the Appendix.
The probability of $\alpha > 90^{\circ}$ reaches its maximum of 56\% at $\theta \approx 0$ (but not exactly equal to $0$).
We find that the kick angle is likely to be within $\sim 30^{\circ}$ of the pre-SN orbital plane so that large BH spin tilts are not unlikely, i.e., $p(\alpha > 90^{\circ})\gtrsim 20\%$.

\section{Discussion}
\label{sec:discuss}

Here we take a step back and interpret the spin measurements of GW200115 (as well as GW200105) in different formation scenarios.
This is necessary because of the lack of conclusive evidence for BH spin in GW200115 as pointed out in Section \ref{sec:spin}.

\subsection{The isolated binary evolution channel: the first-formed compact object is a BH}
\label{sec:discuss1}

This is the scenario explored in Section \ref{sec:constrain}, which is expected to dominate the canonical formation channel \citep[e.g.,][]{Floor_NSBH21}.
First, we note that in some population synthesis studies \citep[e.g.,][]{Debatri21_nsbh}, the term ``BH-NS" is adopted to specifically separate from ``NS-BH" which refers to the scenario where the NS is formed first as a result of mass-ratio inversion during mass-transfer episodes; we discuss the latter in subsection \ref{sec:discuss2}.
For simplicity, we adopt the term NS-BH throughout this paper.

The key assumption made in Section \ref{sec:constrain} is that the BH is rapidly spinning.
The prior constraint on BH spins being discussed here is rather limited.
\citet{Debatri21_nsbh} assumed the BH to possess zero spin\footnote{See also \citet{MandelFragos20} for a summary of theoretical expectations of BH spin magnitudes in BBH mergers.}, motivated by the model of \citet{FullerMa19}.
The argument is that the majority of angular momentum of massive stars is lost as star envelopes are stripped off, if one assumes efficient transport of angular moment inside the stars.
However, that is not expected to be the final word for the following reasons.

First, while the \citet{FullerMa19} model seems to explain reasonably well the low effective spins\footnote{The effective spin is defined to be a mass-weighted spin projection along the orbital angular momentum vector. A small effective spin could mean a small spin or a large spin that is directed toward the orbital plane. Misaligned spins cause spin precession effects which are difficult to measure.} of ten binary BH (BBH) mergers detected in the first two LIGO/Virgo observing runs \citep{BBHpopLVC_O1O2}, it remains challenging to draw a firm conclusion about the distribution of BH spin magnitudes even with an increased sample of 44 BBH observations \citep{GWTC2_pop}; see also \citet{Roulet21_BBHspin} and \citet{Shanika21BBHspin} based on different parameterized models.

Second, all three BHs with available spin estimates in high-mass X-ray binaries \citep{LiuJF08,Orosz09,GouLJ14} are found to be spinning at near-maximum rates ($\chi \sim 1$), which is difficult to explain through accretion because of the short lifetimes of their high-mass companions; see \citet{Miller15review} for a review and \citet{QinBHspin19} for a specific model to explain the large BH spins in these systems.
Of particular relevance among these three systems is Cygnus X-1, which used to be considered as a potential progenitor of NS-BH systems \citep{Belczy11CX-1}.
Recently, \citet{Coen21CX-1} adopted new mass estimates of the BH and its companion star, both higher than previously thought following the revised distance measurement \citep{JamesMJ21_CX1}; they found that Cygnus X-1 may evolve into a BBH system but unlikely to merge within a Hubble time.
However, while bearing in mind the difficulties in predicting the future of massive binaries \citep{Belczynski2108}, it is understood that a merging BBH or even NS-BH system is not ruled out for Cygnus X-1 \citep{Coen21CX-1}.

Therefore, perhaps the strongest argument for a small BH spin in NS-BH binaries comes from the observation of GW200105, for which the spin magnitude is constrained to be below 0.23 with 90\% confidence.

\subsection{The isolated binary evolution channel: the first-formed compact object is a NS}
\label{sec:discuss2}

Based on current understanding of massive binary evolution, this is a less likely scenario because it requires highly efficient mass transfer so that the initially more massive star in the binary lose so much mass that it collapses into a NS instead of a BH \citep[e.g.,][]{Sipior04PSR-BH}.
Through a series of population syntheses, \citet{Floor21a} found that this subchannel is at least a factor of four less likely than that discussed in subsection \ref{sec:discuss1}.
Nevertheless, we briefly comment on this scenario in light of the BH spin measurements.

Motivated by the model of \citet{QinY18BHspin} which was developed for BBH mergers, \citet{Debatri21_nsbh} assumed that the spin of the secondly-formed BH is linearly correlated with the orbital period prior to the second SN.
In this model, BHs formed at shorter pre-SN orbital periods spin faster because their progenitor helium stars are spun up by tidal interactions; the spin magnitudes can be anywhere from zero to maximum.

If GW200115 is formed from this subchannel and the BH is spinning rapidly, the large spin tilt angle would be difficult to account for with BH natal kicks, under similar assumptions about the pre-SN binary configuration made in Section \ref{sec:constrain}; see \citet{Fragos10spin} for a population synthesis study of BH X-ray binaries, and \citet{Gompertz21} for a recent study specifically on NS-BH discoveries.
It is worth mentioning that PSR J0737$-$3039B in the Double Pulsar system, which is the secondly-formed NS, has a spin tilt angle of $130^{\circ}$ \citep{DPSRspin_precess08}. However, the off-center kick model, proposed to explain the pulsar spin-tilt measurement by \citet{Farr11_DPSRspin}, has no explanatory power for spin-orbit misalignments of BHs in X-ray binaries \citep{Salvesen20}.

\subsection{Other possibilities}
\label{sec:discuss3}
The case that GW200115 contains either a non-spinning BH or a fast-spinning BH with a large spin tilt angle would have been consistent with a dynamical origin in dense stellar environments such as globular clusters \citep[e.g.,][]{Rodriguez16_spin}.
However, the dynamical merger rate predictions are too low \citep{Clausen13,Arca20,Ye20rate} to match the measured value in \citet{LVKnsbh21}.

In \citet{LVKnsbh21}, another two formation channels are listed as being capable of individually accounting for the measured NS-BH merger rate: a) the young star cluster channel, and b) the active galactic nuclei disk channel.
It is unknown what expectations on the BH spin magnitudes and tilt angles are for these two channels.
For channel a), the BH spin characteristics might resemble either the isolated binary evolution channel or the dynamical formation channel, depending on whether the BH and NS are formed in a primordial binary or paired up through dynamical interactions \citep{Rastello20_nsbh}.
For channel b), the predicted NS-BH merger rate can only match the measured rate if a significant fraction of LIGO/Virgo BBHs are also formed in this channel \citep{McKernan20_AGNchannel}.

\section{Conclusions}
\label{sec:conclude}
The recent discovery of gravitational waves from two merging NS-BH binaries offers exciting opportunities to study the formation of this new subclass of compact binary systems.
The most interesting feature of these two systems is that the BH spin in GW200115 appears to be negatively aligned with the binary orbital angular momentum vector.
We reassess the observational evidence for BH spin in this gravitational-wave event and find that it is inconclusive.

Assuming the BH in GW200115 is indeed spinning rapidly, we show that its spin-orbit misalignment angle ($\alpha$) implies tight constraints on NS natal kicks in the isolated binary evolution scenario.
We find that $\alpha > 90^{\circ}$ in GW200115-like systems would indicate not only  a large NS natal kick velocity $\sim 600 \, \text{km/s}$ but also a rather restricted kick angle that is $\sim 30^{\circ}$ within the pre-SN orbital plane.
This highlights the power of gravitational-wave observations to potentially probe supernova physics.

It is possible that the BH in GW200115 is slowly spinning (similar to GW200105) and thus no constraint can be derived from the apparent spin-orbit misalignment.
This would indicate that the angular momentum transport within massive stars is highly efficient.
This also highlights the need to develop realistic spin models for gravitational-wave analysis, which presumably require detailed numerical simulations to compute BH spin magnitudes in NS-BH binaries.
A realistic BH spin model will also facilitate the astrophysical inference with a population of $\mathcal{O}(\text{dozens})$ of NS-BH mergers that will be detected over the next five years.

\begin{acknowledgments}
I thank Ilya Mandel for insightful conversations and the anonymous referee for useful comments.
\end{acknowledgments}

\appendix

\begin{figure}[ht]
\begin{center}
  \includegraphics[width=0.5\textwidth]{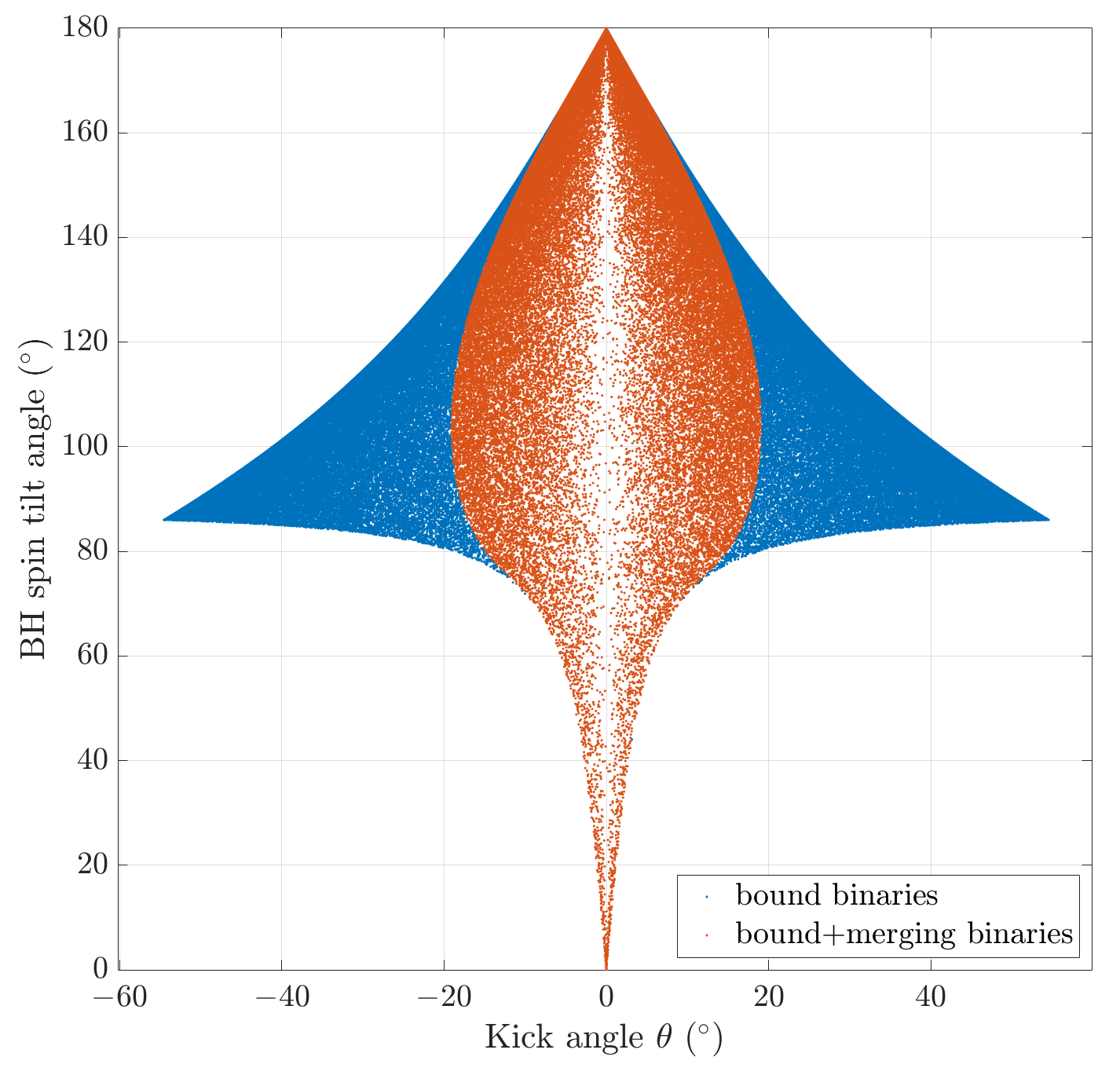}
  \caption{Constraints on the NS natal kick angle $\theta$ and the BH spin tilt angle from the requirement that the post-SN binary (1) remains bound and (2) is capable of merging within a Hubble time. We perform the calculation for $10^{6}$ binaries, each with a random kick drawn from an isotropic distribution, while all other parameters are kept identical. Blue dots indicate binaries that remain bound, while orange dots are those remain bound \textit{and} are capable of merging within a Hubble time. \label{fig:kick_tilt}}
\end{center}
\end{figure}

To further explain the result shown in Figure \ref{fig:kick_angle}, we perform the calculation for one specific binary configuration: $(m_{1}, m_{2})=(5.9, 1.4) M_{\odot}$, $m_{\rm CO}=3 M_{\odot}$, $a_{0} =20 R_{\odot}$.
(The pre-SN orbital velocity is $291 \, \text{km/s}$.)
Based on our kick prescription, the NS natal kick velocity is $457 \, \text{km/s}$.
With random kicks applied to a million such binaries, Figure \ref{fig:kick_tilt} shows the resultant BH spin tilt angle versus the kick angle $\theta$: binaries that remain bound are in blue, while those that remain bound and are merging are in orange.
One can see that 1) the ``merging" requirement leads to a preference for small $\theta$, and 2) to get a non-zero spin tilt angle, the kick angle $\theta$ can be close to zero but not exactly zero.

\bibliography{ref}

\end{document}